# Femtosecond LIPSS on indium-tin-oxide thin films at IR wavelengths

**Balázs Bánhegyi** [1,2*], **László Péter** [1], **Péter Dombi** [1,3], **Zsuzsanna Pápa** [1,3]

[1]*Wigner Research Centre for Physics, Konkoly-Thege Miklós út 29-33, 1121 Budapest, Hungary*
[2]*Department of Atomic Physics, Budapest University of Technology and Economics, Budafoki út 8, 1111 Budapest, Hungary*
[3]*ELI-ALPS, ELI-HU Nonprofit Kft., Wolfgang Sandner utca 3, 6728 Szeged, Hungary*
*\*Corresponding author: banhegyi.balazs@wigner.hu*

We investigated laser-induced periodic surface structures (LIPSS) generated on indium-tin-oxide (ITO) thin films with femtosecond laser pulses in the infrared region. Using pulses between 1.6 and 2.4 µm central wavelength, we observed robust LIPSS morphologies with a periodicity close to λ/10. Supporting finite-difference time-domain calculations suggest that the surface forms are rooted in the field localization in the surface pits leading to a periodically increased absorption of the laser pulse energy that creates the observed periodic structures.

## 1. INTRODUCTION

Indium-tin-oxide (ITO) is one of the most widely used conducting oxides due to its high electrical conductivity and broad spectral transmission from the visible to the infrared range. Deposited in the form of thin surface film, it can serve as a transparent electrode or conductive layer in nanoplasmonic experiments [1], electrochemical systems [2], photovoltaic devices [3-5] or even biosensors for virus detection [6]. Previously it was shown that the surface of an ITO thin film can be patterned with ultrashort laser pulses and certain characteristics of the treated areas can be altered. Many of these comparative studies investigated the laser-induced periodic surface structure (LIPSS) generation both experimentally and theoretically.

In general, LIPSS formations can be divided into two categories based on their periodicity compared to the laser wavelength. i) In the case of low-spatial-frequency LIPSS (LSFL), the laser wavelength and the periodicity of the generated structures are on the same order of magnitude and the orientation of the structures can be either parallel or perpendicular to the laser polarization. ii) The periodicity of the high-spatial-frequency LIPSS (HSFL) is considerably smaller than the applied laser wavelength, and these structures are usually perpendicular to the laser polarization [7]. The formation mechanism of LIPSS generated by ultrafast laser pulses with different wavelengths and pulse durations has been extensively examined in recent years. It is generally accepted that the perpendicular LSFL generation on materials with large electron concentration is caused by the interaction of the incident laser field and the electromagnetic wave on the surface scattered by the initially rough surface [8-10]. The interference of these waves distorts the spatial field distribution along the surface, thus forming a periodic intensity pattern that leads to the corresponding ablation. LSFL with parallel orientation is mainly observed on dielectrics due to a specific nonpropagating electromagnetic mode close to the rough surface (radiation remnants) [7, 11]. On the other hand, explaining the emergence of HSFL patterns and their periodicities is more challenging. Previously it was proposed that the interference of the incident laser radiation with the excited sub-wavelength surface plasmon polaritons could be responsible for the formation of periodic sub-wavelength ripples [12]. Another study presented numerical simulations based on Maxwell equations coupled with laser-induced electron plasma kinetics that addressed the organization of high-frequency, sub-wavelength nanogratings on the surface and in the volume of transparent dielectrics with orientation orthogonal to the laser polarization [13, 14]. Furthermore, coupled electromagnetic and compressible Navier-Stokes simulations showed that the transverse temperature gradients triggered by non-radiative optical response of surface topography could be the origin of high spatial frequency periodic structures on metal surfaces, with dimensions down to λ/15 [15].

Focusing onto the patterning of ITO thin films, previous works reported LIPSS generation in a wide range of experimental conditions regarding the laser wavelengths, pulse duration and repetition rate. Using pulses with central wavelengths between 532 and 1032 nm both in the fs and ps regime resulted in HSFL formations with periodicities between 150 and 350 nm with perpendicular orientation [16-20]. Overall, the reported periodicities and the resulting morphologies are similar and show minor dependence on the incident laser wavelength in the previously investigated visible and near-infrared range. This suggests that the formation of the HSFL morphologies and the resulted periodicities could be in a strong connection to some fundamental parameter of the ITO itself. To verify this hypothesis, we set out to conduct a comparative study that explores the generated LIPSS formations produced by femtosecond laser pulses having different central wavelengths in the infrared regime (1.6, 2.0 and 2.4 μm). The resulting periodic structures on the surface of the ITO and the underlying glass are compared to our finite difference time domain (FDTD) simulations to reveal the processes of nanostructure formation and its evolution.

## 2. EXPERIMENTAL SETUP AND METHODS

In our experiments, we used a regenerative Ti:sapphire amplifier (Coherent Legend Elite) seeded with a home-built oscillator. The 800 nm, 50 fs output pulses with 10 kHz repetition rate were directed into an optical parametric amplifier (OPA) with a 70-fs output pulse length, and controllable output wavelength between 300 nm and 15 μm. The sample was a 150 nm thick ITO layer deposited onto a 1 mm fused silica substrate. These infrared pulses were focused by back-illumination to the ITO-glass interface by a 200 mm focal length $CaF_2$ lens, resulting a focal spot with 260 μm diameter. All LIPSS generation experiments were conducted in ambient air environment. Pulse energies were controlled by a set of discrete neutral density filters. In the investigated wavelength range, the ITO has an increasing absorption together with the decrease of the transmission (Fig. 1 (a)). IR transmission of the sample was measured with a PerkinElmer Lambda 1050 spectrophotometer. Optical properties of the ITO layer in the investigated spectral range was deduced using the refractive index and extinction coefficient datasets of Cleary et al. [21], which were modified slightly to fit it to the measured transmission spectrum (Fig. 1 (b)).

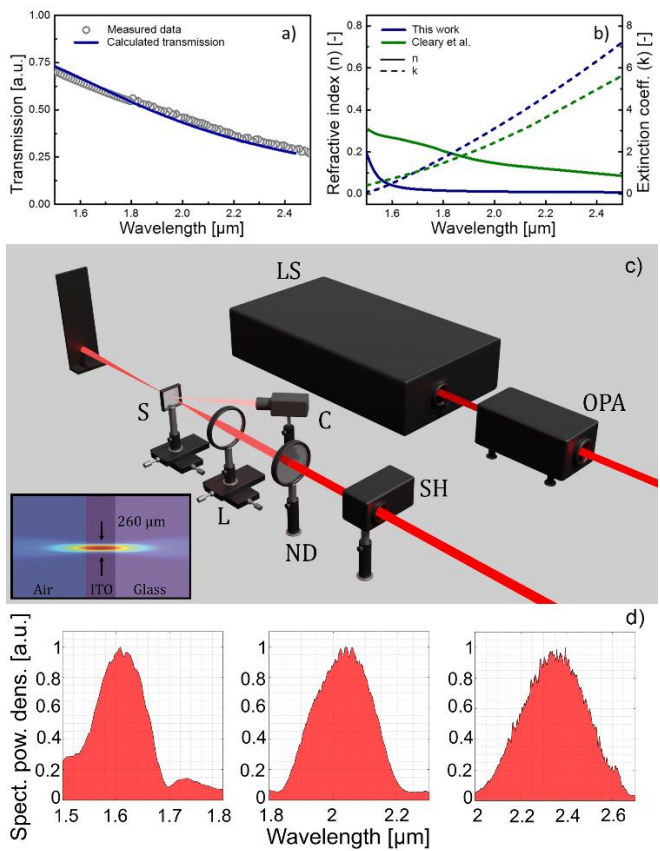

Fig. 1. a) Measured and calculated transmission spectrum of the ITO layer. b) Optical properties of the ITO layer. Slight modification of the literature datasets were carried out in order to describe properly the measured transmission spectrum. c) Experimental setup with femtosecond light source (LS), tunable optical parametric amplifier (OPA), programmable shutter (SH), neutral density filter (ND), focusing lens (L), ITO covered glass sample (S) and CCD camera (C). Inset: back-illuminated focusing geometry with normalized electric field distribution. d) Infrared spectra of the at the applied wavelengths.

Considering the increasing absorption of the ITO, the absorbed pulse energy is higher at longer wavelengths, which was compensated by using less energetic pulses for longer wavelengths. Illumination times between 0.1 s and 5 s were used and the duration was set by a controllable shutter. Due to the subsequent nonlinear processes in the OPA, negligibly low energy pulses with longer duration in the visible range were also present at the output besides the infrared pulses used to generate the LIPSSs. These pulses did not contribute to the periodic structure generation processes but were used to monitor the LIPSS formation by detecting the scattered light from the modified surface with a CCD camera (Fig. 1 (c)). The laser-treated areas and the formed LIPSSs were inspected with a scanning electron microscope (TESCAN MIRA3 SEM). To resolve the surface morphology, the in-beam secondary electron detector was chosen with 5 kV accelerating voltage. The images were acquired via line accumulation with 4 mm working distance and a magnification between 2400 and 160 000. To gather information about the composition and to examine the amount of ITO ablated from the surface, we also performed element analysis in the laser-treated areas. Before the LIPSS experiments, the surface of the original, untreated ITO layers was also investigated with a PSIA XE-100 atomic force microscope in dynamic, non-contact mode over 5 μm × 5 μm and 20 μm × 20 μm scanning areas. The average roughness was determined by the manufacturer's (XEI) software.

## 3. RESULTS AND DISCUSSION

### A. Exposition for 0.1 s

Fig. 2. shows SEM images of the surface morphologies resulting after 0.1 s exposition with pulse energies between 9 and 15 μJ.

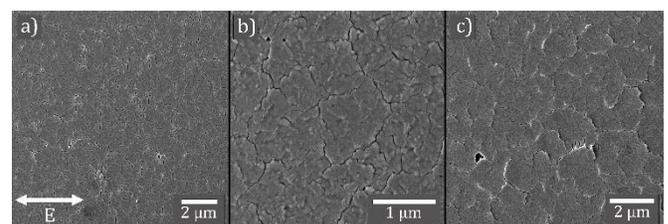

Fig. 2. SEM images of the generated LIPSSs using pulses with a) 1.6 μm wavelength and $3.8 \times 10^{11}$ W/cm$^2$ peak intensity, b) 2.0 μm wavelength and $3.8 \times 10^{11}$ W/cm$^2$ peak intensity and c) 2.4 μm wavelength and $2 \times 10^{11}$ W/cm$^2$ peak intensity after 0.1 s of illumination.

The 0.1 s illumination with infrared pulses resulted in cracks on the ITO surface in the case of all three wavelengths.

The orientation of these cracks does not show any dependence on the laser polarization and does not follow the initial ITO morphology. This suggests that the origin of these fractures is the heat stress caused by the rapid heating of the laser-treated area due to the absorbed energy of the laser pulses. This is also supported by similar observations of Steinecke et al. when annealing ITO at high temperatures [22]. Fig. 2(a-c) also show the presence of few nanometers sized craters where the ITO got completely removed from the glass substrate. These can be explained by the following: As the fractures form on the surface, regions, where the initial ITO cohesion is weaker, get separated from the surrounding material and get rejected from the surface due to the buildup of local electrostatic charge. It can be also observed that apart from these fractures and craters, the ITO surface morphology remained intact, there are no signs of any melting and resolidification processes. These changes of the ITO surface pave the way for more prominent surface modifications of subsequent pulses.

**B. Exposition for 1 s**

SEM images of the generated LIPSS formations in the case of 1 s exposition with pulse energies between 9 and 15 µJ are shown in Fig. 3.

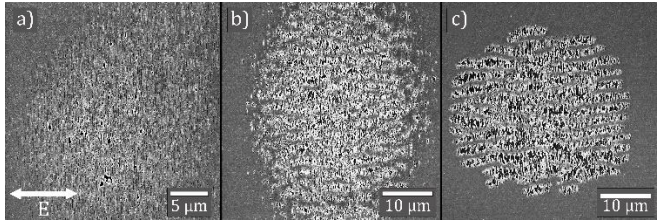

Fig. 3. SEM images of the generated LIPSS structures using pulses with a) 1.6 µm wavelength and $3.8 \times 10^{11}$ W/cm² peak intensity, b) 2.0 µm wavelength and $3.8 \times 10^{11}$ W/cm² peak intensity and c) 2.4 µm wavelength and $2 \times 10^{11}$ W/cm² peak intensity after 1 s of illumination.

As shown in Fig. 3(a) using 1.6 µm pulses, cracks appear on the surface of the ITO with an orientation perpendicular to the laser polarization. Close to the center of the focal spot in the highest local intensity region several randomly arranged craters are present with sizes between 150 and 500 nm. In the case of 2 µm and 2.4 µm pulses, where the absorption of the ITO is higher, periodic patterns appear forming HSFLs perpendicular, and LSFLs parallel to the laser polarization. The LSFLs created by 2-µm pulses are irregular, and they become more distinct in the case of 2.4 µm illumination. The periodicity of these structures is close to 2 µm based on the SEM images, without a clear wavelength-dependence. Around the structured area created by 2 µm pulses, a transition zone can be observed where the lower local intensity creates nanosized craters similar to the ones generated by the 1.6 µm pulses. This transition zone disappears in the case of 2.4 µm illumination.

1. HSFL structures

Focusing first on the HSFL structures, their properties can be analyzed in more detail by taking a closer look the illuminated spots. The generated HSFL formations for all three wavelengths are shown in Figs. 4(a-c) with higher magnification. Fig. 4(d) shows the deduced periodicities. The periodicity of the HSFLs was calculated by applying a 2-dimensional fast Fourier transform (FFT) to the SEM images. Based on the FFT inverse-space spectra, the periodicity of the HSFLs is between 150 and 300 nm, with a slight increase for larger wavelengths (Fig. 4(d)).

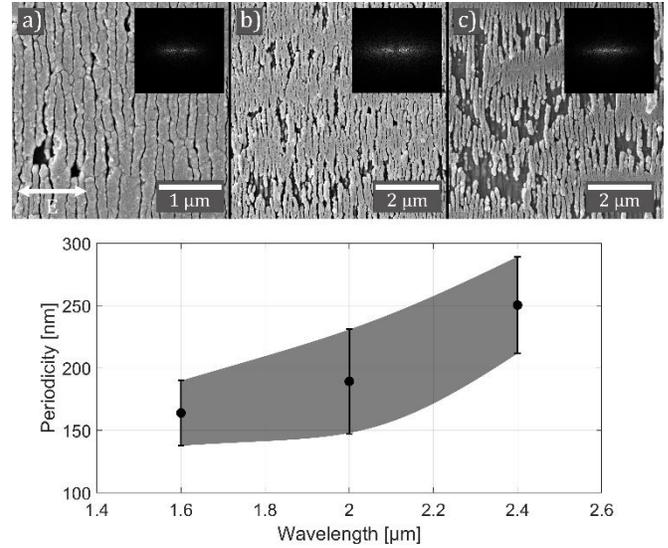

Fig. 4. Detailed SEM images of the HSFL formations in case of a) 1.6 µm and $3.8 \times 10^{11}$ W/cm², b) 2.0 µm and $2.3 \times 10^{11}$ W/cm², c) 2.4 µm and $2 \times 10^{11}$ W/cm² peak intensity pulses. Inset figures show the FFT invers-space spectra of the SEM images unveiling real space periodicity. d) Period length values with error bars calculated based on the distribution of FFT spectra.

The periodicity of the HSFLs was calculated by applying a 2-dimensional fast Fourier transform (FFT) to the SEM images. Based on the FFT inverse-space spectra, the periodicity of the HSFLs is between 150 and 300 nm, with a slight increase for larger wavelengths (Fig. 4(d)). These structures having period lengths much smaller than the wavelength and perpendicular orientation, can be discussed in the frame of electromagnetic simulations, where they are often called r-type LIPSS [23-25]. This means that they are connected to the initial roughness of the surface. To check this explanation, we took into account the roughness of the ITO layer, which was studied with an AFM before any LIPSS experiment took place. Although the average roughness is not very high (rms roughness < 5 nm), the microroughness of the surface has a similar characteristic length to the period of the observed LIPSS structure. Fig. 5(a) shows a 3D visualization of the surface, while in b) two lineouts of the scanned area are presented.

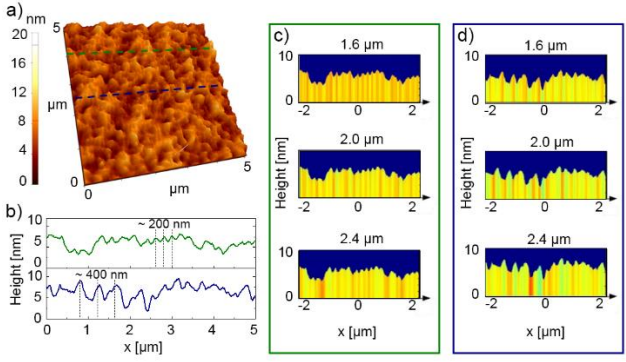

Fig. 5. a) 3D visualization of the surface based on AFM data. b) Lineouts of the scanned area. Characteristic period of the initial surface structures is 200-400 nm. c) and d) Normalized values of the absorbed power inside the ITO layer along a line section for wavelengths 1.6 μm, 2 μm and 2.4 μm, respectively. Note that the line sections are the same as in panel b).

Based on the electromagnetic theory, the interaction of the incident light with the initial structure of the irradiated material can lead to field localization in the pits of the rough surface. These pits of the surface may act as seeds for the more prominent structural changes or even material removal [26]. To investigate the possible field localization, we calculated the distribution of the electric field inside the ITO with Lumerical FDTD Solutions software. For describing ITO, we used the optical properties shown in Fig. 1 (b), while fused silica substrate was modeled based on the values of [27]. The source properties (bandwidth, temporal length, polarization) were set to be the same as of the laser pulses during the experiment. We applied backside illumination in the simulations in order to reproduce the experimental conditions. For being able to monitor the effect of the surface roughness, AFM images taken from the surface of the ITO layer were directly introduced to the simulations. According to the manufacturer of the ITO coated samples (Präzisions Glas & Optik GmbH), the rms roughness of fused silica is below 1 nm, therefore, the roughness of the fused silica substrate was not included into our simulations. Volume of the FDTD unit cell was 5 μm × 5 μm × 1 μm and the spatial resolution near the surface of the ITO was set to 1 nm. Distribution of the electric field was monitored inside the ITO along given lineouts. For visualizing the possible positions of field localization and material removal, it is worth to plot absorbed energy values being proportional with the electric field square:

$$P_{abs} = -0.5\omega|\vec{E}|^2 Im(\varepsilon) \quad (1)$$

where ω is the angular frequency of the illumination, E is the electric field and ε is the dielectric function of the ITO at the given angular frequency value. Fig. 5(c-d) show the normalized values of the absorbed power inside the ITO layer along two line sections.

Positions of regions encountering the maximum absorbed energy correspond to the pits on the ITO surface. These patterns of maximum absorbed energy define precisely the thermally affected zones, where the material modification takes place with the highest probability [16]. Absorbed energy distributions also hint, that larger wavelengths are more effectively absorbed at larger structures, which may result in the slight increase in the observed period length. Although this simple interpretation does not account for multipulse effects, it still predicts that during our experiments, it is most likely the initial roughness of the ITO layer, and the field localization in the pits of this rough surface that governs the formation of the HSFL structures.

We can get another indication of the roughness-based origin of the observed structures if we plot the field distribution maps recorded in the top 10 nm depth of the layer. These averaged electric field patterns belonging to the applied wavelengths (1.6 μm, 2.0 μm, and 2.4 μm) are plotted in Fig. 6 (a-c) (values are normalized). These figures can indicate the formation of the HSFL structures since the spatial domains exhibiting the largest local fields correspond to the structures of the rough surface. Furthermore, if the electric field pattern is Fourier-transformed, the vertical orientation of the high-intensity domains is clearly visible (note the bright areas located along the horizontal axis in the insets of Fig. 6(a-c)). The periodicity of the brightest vertical domains is the smallest for 1.6 um, varying between 180 and 400 nm, while for the other two wavelength it is slightly larger (190-430 nm for 1.6 μm and 200-500 nm for 2.4 μm) matching nicely to the slight wavelength dependence of the periodicity of the LSFL structures (c. f. Fig. 4(d)). Furthermore, if we compare our data with the literature results of ITO patterning, we can experience that despite the diverse experimental conditions of previous studies and our investigations, the appearing HSFLs have very similar properties regarding both the orientation and the periodicity. This also suggests that it is more likely the ITO itself (and its properties) that determines the structure formation and the wavelength/duration of the applied laser pulses has less impact.

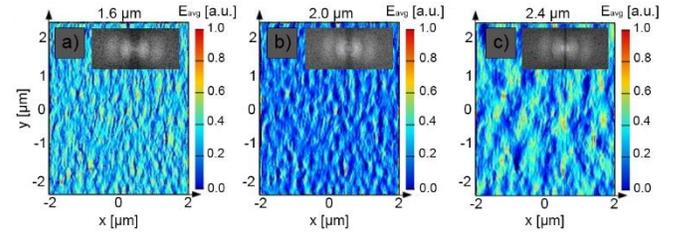

Fig. 6. a) Normalized average electric field distribution along the rough surface of the ITO layer using a) 1.6 μm, b) 2 μm and c) 2.4 μm illumination wavelength. Insets: FFT-spectra of the intensity distribution show the vertical periodicity clearly with characteristic length ranging between 180 and 500 nm.

2. LSFL structures

Concentrating onto the LSFL structures appearing at 2 and 2.4 μm wavelength, it is important to mention that the parallel orientation of LSFL structures is not characteristic of conductive layers. To gather more information about the formation of these surface structures, we performed element analysis around the laser-treated areas. A 10 kV electron accelerating voltage was used for the sake of achieving a sufficient X-ray yield for all EDS lines (that all lie in the < 4 keV regime) and to keep the incidence depth of the electron beam relatively low. The typical excitation volume under the above-mentioned circumstances is around 0.2 μm³, which means that the composition information is averaged over this volume. Taking into account the limit above, we performed a line scan on the structured area generated by illuminating the sample 1 s with 2.4 μm wavelength, 2 × $10^{11}$ W/cm$^2$ peak intensity pulses.

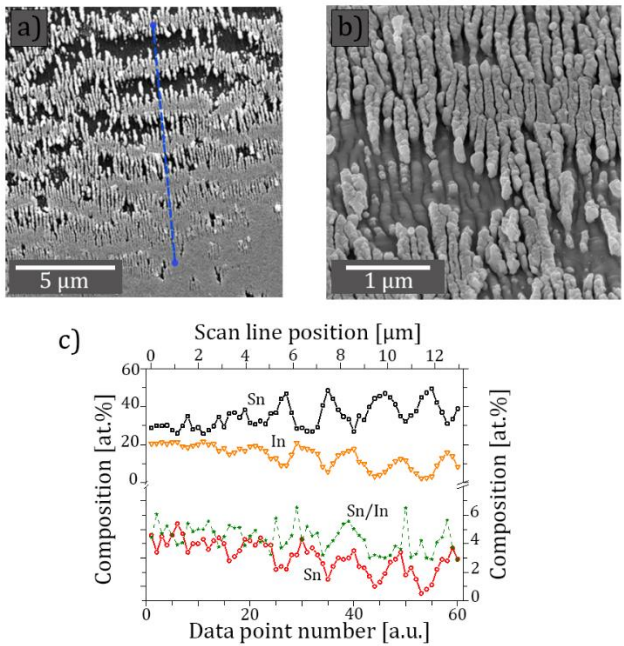

Fig. 7. a) Element analysis results obtained by a line scan on the laser treated area. The dashed blue line shows the line path. b) Tilted SEM image of the HSFL formations of the ITO and the structured underlying glass generated with 2.4-μm wavelength, 2.3 × $10^{11}$ W/cm$^2$ peak intensity pulses with 1 s illumination. c) Local composition of the different components on the surface along the scan line.

Fig. 7. shows the line scan path and the obtained relative density of elements along the scan based on the detected electrons originating from the listed shells. Approaching the center of the focal spot, the Si concentration starts to oscillate as the ITO becomes periodically ablated and the underlying glass becomes uncoated. At the same time, in an opposite phase, the In and Sn concentrations also oscillate showing the same effect. This clearly shows that LSFL structures form through this periodic ablation of the ITO layer. This means that there has to be a periodically larger local field that promotes the ablation. The physical origin of this periodic ablation is still an open question. The orientation and periodicity of the LSFL can be related to non-metallic surfaces, which hints that these structures may origin from a periodic intensity pattern developing on the substrate surface [11]. However, since the optical properties of fused silica are almost constant between 1.6 and 2.4 μm, it is expected that LSFL should appear at all illumination wavelengths, which is not the case.

When investigating tilted SEM images with higher magnification, it can be noticed that the glass surface is also periodically structured, as shown in Fig. 6(c). In areas where the ITO became ablated, the underlying glass substrate reveals its structured surface. The subwavelength periodicity of this pattern is comparable to the periodicity of the ITO. This patterning of the substrate is very similar to glass imprinting by ITO and ZnSe layers at low fluences compared to the ablation threshold of the substrate [12, 19, 20].

### C. Exposition for 5 s

Longer expositions with higher pulse energies result in the ablation of the ITO in the center of the focal spot. Fig. 8 shows the laser treated area after 5 s exposition of 2-μm pulses with 15 μJ pulse energy.

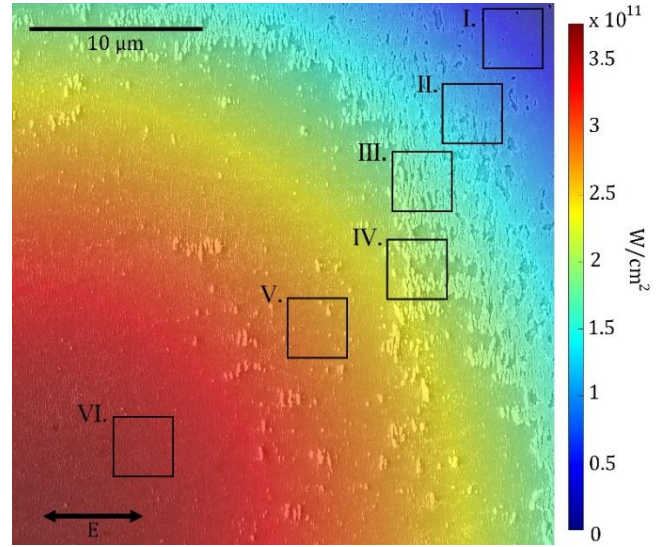

Fig. 8. SEM image of the laser treated area color-mapped with the local peak intensity using 2-μm wavelength, 3.8 × 1011 W/cm2 peak intensity pulses with 5 s illumination. For a description of areas I-VI, see manuscript Table 1.

The local intensity variations due to the Gaussian distribution enable to identify threshold intensity values for the different surface morphologies. Typical morphologies around the focal spot can be classified as follows: I. At lower peak intensities the ITO starts to crack due to the thermal stress caused by higher thermal gradient. II. At higher peak intensities, smaller craters appear, and a primary periodicity perpendicular to the electric field can be seen comparable to the one in Fig. 3(a). III. Increasing the peak intensity these vertical stripes start to separate from each other creating HSFL with a periodicity of 150 – 300 nm perpendicular to the laser polarization. IV. Closer to the focal spot, the secondary LSFL is formed with a periodicity around 2 μm, matching the laser wavelength. V. Close to the center of the focal spot the ITO is almost fully ablated from the glass substrate. The remaining ITO is present in the form of melted droplets with varying sizes. VI. In the center of the focal spot, where the local intensity is the highest, there is no ITO left, only the underlying glass substrate is present. All these surface morphologies and the corresponding threshold peak intensities are summarized in Table 1.

**Table 1. Different morphologies formed on the ITO and glass substrate surface and the corresponding threshold peak intensities.**

| Area in Fig. 5 | Peak intensity [W/cm$^2$] | Surface morphology |
|---|---|---|
| I. | $2.1 \times 10^{10}$ | Cracks appear on the ITO surface due to thermal shock. |
| II. | $7.1 \times 10^{10}$ | Nanometer sized craters and vertical periodic splinters. |
| III. | $1.2 \times 10^{11}$ | Separated periodic stripes with 150 – 300 nm periodicity. |
| IV. | $1.6 \times 10^{11}$ | Secondary, horizontal structure with 2 μm periodicity. |
| V. | $2.6 \times 10^{11}$ | Partly ablated ITO with melted droplets. |
| VI. | $3.2 \times 10^{11}$ | Ablated ITO, periodically structured glass surface. |

This representation also supports that HSFL appears first on the ITO surface, and the ablation rooted LSFL forms only at higher intensities, as a next step.

## 4. CONCLUSIONS

We performed experiments in which we used femtosecond pulses in the infrared region with three different wavelengths to generate LIPSS on an ITO thin film. We observed primary HSFLs with periodicity between 150 and 300 nm with an orientation perpendicular to the laser polarization. The much smaller periodicity compared to the wavelength can be interpreted as a roughness-dependent structure. FDTD simulations based on AFM measurements were conducted, that confirmed our hypothesis about the origin of the LIPSS formations. We concluded that the initial surface roughness of the ITO layer, and the field localization in the pits lead to a periodic absorption of the laser pulse energy that creates the periodic structures and even periodic material ablation from the substrate. We also obtained LSFLs parallel to the laser polarization with 2 μm periodicity, which we investigated with element analysis unfolding the periodic concentration of the surface material components.

**Funding.**

**Acknowledgments.**

**Disclosures.** The authors declare no conflicts of interest.

**Data availability.** No data were generated or analyzed in the presented research.

## References

1. Kanehara, M., Koike, H., Yoshinaga, T. and Teranishi, T. Indium Tin Oxide Nanoparticles with Compositionally Tunable Surface Plasmon Resonance Frequencies in the Near-IR Region. *Journal of the American Chemical Society* **131**, 17736-17737 (2009) https://doi.org/10.1021/ja9064415

2. Sakamoto, K., Kuwae, H., Kobayashi, N., Nobori, A., Shoji, S. and Mizuno, J. Highly flexible transparent electrodes based on mesh-patterned rigid indium tin oxide. *Scientific Reports* **8**, 2825 (2018) https://doi.org/10.1038/s41598-018-20978-x

3. Jahng, W. S. and Francis, A. H. Is indium tin oxide a suitable electrode in organic solar cells? Photovoltaic properties of interfaces in organic p/n junction photodiodes. *Applied Physics Letters* **88**, 093504 (2006) https://doi.org/10.1063/1.2180881

4. Al-Ibrahim, M., Roth, H. K. and Sensfuss, S. Efficient large-area polymer solar cells on flexible substrates. *Applied Physics Letters* **85**, 1481–1483 (2004). http://doi.org/10.1063/1.1787158

5. H. Liu, and R. Sun, Laminated active matrix organic light-emitting devices. *Applied Physics Letters* **92** 063304 (2008). https://doi.org/10.1063/1.2844854

6. Guo, D., Zhuo, M., Zhang, X., Xu, C., Jiang, J., Gao, F., Wan, Q., Li, Q. and Wang, T. Indium-tin-oxide thin film transistor biosensors for label-free detection of avian influenza virus H5N1. *Analytica Chimica Acta* **773**, 83–88 (2013) https://doi.org/10.1016/j.aca.2013.02.019

7. Bonse, J., Höhm, S., Kirner, S. V., Rosenfeld, A., and Krüger, J. Laser-Induced Periodic Surface Structures - A Scientific Evergreen. IEEE JOURNAL OF SELECTED TOPICS IN QUANTUM ELECTRONICS **23**, 9000615 (2017)

8. Young, J. F., Preston, J. S., van Driel, H. M. and Sipe, J. E. Laser-induced periodic surface structure. II. Experiments on Ge, Si, Al, and brass. *Physical Review B* **28**, 1155-1172 (1983) http://doi.org/10.1103/PhysRevB.27.1155

9. Dufft, D., Rosenfeld, A., Das, S., K., Grunwald, R., and Bonse, J., Femtosecond laser-induced periodic surface structures revisited: A comparative study on ZnO. *Journal of Applied Physics* **105**, 034908 (2009) https://doi.org/10.1063/1.3074106

10. Bonch-Bruevich, A. M., Libenson, M., N., Makin, V., S. and Trubaev, V. V. Surface electromagnetic waves in optics. *Optical Engineering* **31**, 718 (1992) https://doi.org/10.1117/12.56133

11. Sipe, J. E., van Driel, H. M., and Young, J. F. Surface electrodynamics: Radiation fields, surface polaritons, and radiation remnants," *Can. J. Phys.* **63** 104–113 (1985)

12. Wang, L., Cao, X.-W., Abid, M.-I., Li, Q.-K., Tian, W.-J., Chen, Q.-D., Juodkazis, S., and Sun, H.-B. Nano-ablation of silica by plasmonic surface wave at low fluence, *Optics Letters* **42**, 4446 (2017) https://doi.org/10.1364/OL.42.004446

13. Rudenko, A., Colombier, J-P., and Itina, T-E., From random inhomogeneities to periodic nanostructures induced in bulk silica by ultrashort laser. *Physical Review B* **93**, 075427 (2016) https://10.1103/PhysRevB.93.075427

14. Déziel, J-L., Dubé, L-J., Messaddeq, S-H., Messaddeq, Y., and Varin, C., Femtosecond self-reconfiguration of laser-induced plasma patterns in dielectrics. *Physical Review B* **97**, 205116 (2018) http://10.1103/PhysRevB.97.205116

15. Rudenko, A., Abou-Saleh, A., Pigeon, F., Mauclair, C., Garrelie, F., Stoian, R., Colombier, J.P. High-frequency periodic patterns driven by non-radiative fields coupled with Marangoni convection instabilities on laser-excited metal surfaces. *Acta Materialia* **194**, 93-105 (2020) https://doi.org/10.1016/j.actamat.2020.04.058

16. Rudenko, A., Mauclair, C., Garrelie, F., Stoian, R. and Colombier, J.P., Light absorption by surface nanoholes and nanobumps. *Applied Surface Science*, **470**, 228-233, (2018) https://doi.org/10.1016/j.apsusc.2018.11.111

17. Straub, M., Afshar, M., Feili, D., Seidel, H. and König, K., Efficient Nanostructure Formation on Silicon Surfaces and in Indium Tin Oxide thin-films by sub-15 fs pulsed near-infrared Laser Light. *Physics Procedia* **12**, 16–23 (2011) https://doi.org/10.1016/j.phpro.2011.03.100

18. Farid, N., Nieto, D. and O'Connor, M.-G., Thin film enabling sub-250 nm nano-ripples on glass by low fluence IR picosecond laser irradiation, *Optics & Laser Technology* **108**, 26–31 (2018) https://doi.org/10.1016/j.optlastec.2018.06.059

19. Chen, L., Cao, K., Liu, J., Jia, T., Li, Y., Zhang, S., Feng, D., Sun, Z., and Qiu, J., Surface birefringence of regular periodic surface structures produced on glass coated with an indium tin oxide film using a low-fluence femtosecond laser through a cylindrical lens. *Optics Express* **28**, 30094-30106 (2020) https://doi.org/10.1364/OE.402037

20. Farid, N., Dasgupta, P. and O'Connor, M.-G., Onset and evolution of laser induced periodic surface structures on indium tin oxide thin films for clean ablation using a repetitively pulsed picosecond laser at low fluence, *Journal of Physics D: Applied Physics* **51**, (2018) https://doi.org/10.1088/1361-6463/aab224

21. Cleary, J. W. Smith, E. M., Leedy, D. K., Grzybowski, G., Guo, I. Optical and electrical properties of ultra-thinindium tin oxide nanofilms on silicon for infrared photonics. *Optical Materials Express* **8**, 1231-1245 (2018) https://doi.org/10.1364/OME.8.001231

22. Steinecke, M., Naran, T., A., Keppler, N., C., Behrens, P., Jensen, L., Jupé, M. and Ristau, D. Electrical and optical properties linked to laser damage behavior in conductive thin film materials. *Optical Materials Express* **11**, 35-47 (2021) https://doi.org/10.1364/OME.410081

23. Skolski, J. Z. P., Römer, G. R. B. E., Vincenc Obona, J., and Huis in 't Veld, A. J. Modeling laser induced periodic surface structures: Finite-difference time-domain feedback simulations. *Journal of Applied Physics* **115**, 103102 (2014) http://dx.doi.org/10.1063/1.4867759

24. Déziel, J-L., Dumont, J., Gagnon, D., Dubé, L. J., Messaddeq, S. and Messaddeq, Y. Toward the formation of crossed laser-induced periodic surface structures. *Journal of Optics* **17**, 075405 (2015) https://doi.org/10.1088/2040-8978/17/7/075405

25. Rudenko, A., Colombier, J-P., Höhm, S., Rosenfeld, A., Krüger, J., Bonse, J. and Itina, T. E. Spontaneous periodic ordering on the surface and in the bulk of dielectrics irradiated by ultrafast laser: a shared electromagnetic origin. *Scientific Reports* **7**, 12306 (2017) http://doi.org/10.1038/s41598-017-12502-4

26. Pan, A., Wang, W., Liu, B., Mei, X., Yang, H. and Zhao, W., Formation of high-spatial-frequency periodic surface structures on indium-tin-oxide films using picosecond laser pulses. *Materials & Design* **121**, 126–135 (2017) https://doi.org/10.1016/j.matdes.2017.02.055

27. Malitson, I. H., Interspecimen comparison of the refractive index of fused silica. *J. Opt. Soc. Am.* **55**, 1205-1208 (1965) https://doi.org/10.1364/JOSA.55.001205